\documentclass[aps,prb,twocolumn,superscriptaddress,showpacs,ams]{revtex4}
\usepackage{graphicx}

\begin{document}

\title{Impurity-enhanced Aharonov-Bohm effect in neutral quantum-ring magnetoexcitons}

\author{L. G.~G.~V. Dias da Silva}
\affiliation{Departamento de F\'{\i}sica, Universidade Federal de
S\~ao Carlos, 13565-905 S\~ao Carlos SP, Brazil}
\affiliation{Department of Physics and Astronomy, Nanoscale and
Quantum Phenomena Institute, Ohio University, Athens, Ohio
45701-2979}

\author{S. E. Ulloa}
\affiliation{Department of Physics and Astronomy, Nanoscale and
Quantum Phenomena Institute, Ohio University, Athens, Ohio
45701-2979}

\author{A. O. Govorov}
\affiliation{Department of Physics and Astronomy, Nanoscale and
Quantum Phenomena Institute, Ohio University, Athens, Ohio
45701-2979}

\date{\today }

\begin{abstract}
We study the role of impurity scattering on the photoluminescence
(PL) emission of polarized magnetoexcitons. We consider systems
where both the electron and hole are confined on a ring structure
(quantum rings) as well as on a type-II quantum dot. Despite their
neutral character, excitons exhibit strong modulation of energy
and oscillator strength in the presence of magnetic fields.
Scattering impurities enhance the PL intensity on otherwise
``dark" magnetic field windows and non-zero PL emission appears
for a wide magnetic field range even at zero temperature. For
higher temperatures, impurity-induced anticrossings on the
excitonic spectrum lead to unexpected peaks and valleys on the PL
intensity as function of magnetic field. Such behavior is absent
on ideal systems and can account for prominent features in recent
experimental results.
\end{abstract}



\maketitle

\newcommand{\be}   {\begin{equation}}
\newcommand{\ee}   {\end{equation}}
\newcommand{\ba}   {\begin{eqnarray}}
\newcommand{\ea}   {\end{eqnarray}}
\newcommand{\imp}   {\mbox{\scriptsize imp}}
\newcommand{\conf}   {\mbox{\scriptsize conf}}
\newcommand{\exct}   {\mbox{\scriptsize exct}}

\section{Introduction}
\label{sec:introduction}

A charged particle moving in a magnetic field acquires a phase
proportional to the applied magnetic flux due to the quantum
interference between different closed paths, giving rise to the
long-studied Aharonov-Bohm effect (ABE). \cite{AharonovBohm} This
effect is specially important if the particle's configuration
space has a ring-like topology, since the interference effects
will create flux-dependent phase {\it differences} of
characteristic size, which can be clearly identified in
experiments. \cite{AharonovBohmExp}

An interesting issue appears when one considers a system of bound
charged particles, forming a {\it composite} neutral object. An
example of such system is a ring-confined exciton, an optically
active electron-hole bound state with experimentally accessible
characteristics.

The optical manifestations of such excitonic Aharonov-Bohm effect
in semiconductor quantum-ring structures has received great
attention from both theoretical
\cite{Chaplik95,Kalameitsev98,Pereyra00,Romer_Raikh00,
Romer_RaikhPSL00,Song_Ulloa,Hu01,Govorov02,UlloaPE02,GovorovPE02,Maslov_Citrin03,Climente03}
and experimental \cite{Lorke00,Bayer03,Ribeiro03} groups. Whereas
one could imagine that no ABE should be expected for neutral
particles, a small but non-vanishing ABE in neutral excitons
confined in one-dimensional rings was theoretically proposed.
\cite{Chaplik95,Romer_Raikh00,Romer_RaikhPSL00} In such systems,
the exciton's finite size allows for an internal polarization of
the positive and negative charges in such a way that the magnetic
flux phases acquired by the electron and hole do not cancel each
other. The ABE amplitude in one-dimensional systems is shown to
depend on the tunnelling amplitude of either electron or hole to
the ``opposite side" of the ring and is exponentially suppressed
when this tunnelling amplitude decreases.
\cite{Romer_Raikh00,Romer_RaikhPSL00}

The ABE is further suppressed when a 2D ``ring stripe" is
considered and both electron and hole are confined by a finite
width potential. \cite{Hu01,Song_Ulloa} However, if {\it
different} confining potentials for the electron and hole are
considered in the structure, a strong effect is expected on the
photoluminescence intensity (PLI).
\cite{Govorov02,UlloaPE02,GovorovPE02} In this case, a net radial
polarization of the exciton naturally exists and the ABE could be
strong enough to be detected by photoluminescence (PL)
experiments. The optical emission is predicted to oscillate as the
ground-state angular momentum changes, generating a series of
``dark" and ``bright" PL emission regions as function of the
magnetic flux. The same underlying principle applies in the study
of the PL response in semiconductor quantum dots with type-II band
alignment. \cite{Kalameitsev98,GovorovPE02,Janssens02} An
enhancement of the ABE is also predicted when an in-plane electric
field is applied. \cite{Maslov_Citrin03}

ABE on charged  \cite{Bayer03} and neutral excitons
\cite{Ribeiro03} has been reported in recent beautiful experiments
and found to be in general agreement with expectations. However,
some of the findings remain not completely understood and
interesting questions remain open. For example, the experimental
field-dependent PL intensity does not fully agree with the
expected result in quantum rings. \cite{Ribeiro03,Evaldo_priv}
Moreover, as the PL signal is collected from an ensemble of
dots/rings over a large area, the role of impurities cannot be
neglected.

In this paper, we discuss the effect of impurity scattering on the
optical properties of quantum rings and study the influence of
perturbative defects in the optical Aharonov-Bohm effect in these
systems. Previous treatments of disorder effects in
magnetoexcitons in quantum-rings have focused attention on the
underlying dynamics of the electron-hole dipole moment.
\cite{Maschke01} We choose a different approach, concentrating on
the effects of symmetry-breaking in the optical emission
intensity.

Our main results can be summarized as follows: even though
impurity scattering leads to mixing of angular momentum states,
signatures of the Aharonov-Bohm effect on a neutral exciton remain
for significant impurity strengths. We further find that finite
temperatures not only produce the monotonic smoothing of spectral
features, but also induce additional characteristics in the PLI
that can be attributed to impurity effects. In fact, this suggests
the use of disorder-induced ABE features as a tool to probe into
the impurity potentials and extract information on the confining
strength and localization length of the hole and electron
wavefunctions. Additionally, we will discuss how our results can
account for recent experimental data on the PLI of type-II QDs.
\cite{Ribeiro03}

The paper is organized as follows: in section \ref{sec:refsys} we
set our system of reference, based on experimentally realized
quantum-ring systems. A general description of the models and main
features is given on section \ref{sec:theomod}. The PL emission,
the core result of the paper, is presented in section
\ref{sec:PLemission}, while our overall conclusions are given in
section \ref{sec:final}.

\section{Quantum ring-like systems}
\label{sec:refsys}


The class of systems we are interested in includes nanoscopic
semiconductor ring-structures with typical carrier confining
radius $R \sim 15-80$nm. The confining radii can be different for
electrons and holes due to sample strain and different carrier
masses, giving rise to a net radial polarization of the
electron-hole pair. A natural assumption is that the radial
confining width $w$ is small ($R/w \gg 1$) for at least one of the
carriers.

\begin{figure}[h!]
\includegraphics[height=0.43\columnwidth,width=0.49\columnwidth]{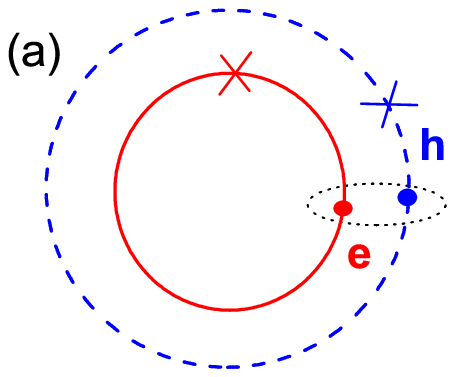}
\includegraphics[height=0.43\columnwidth,width=0.49\columnwidth]{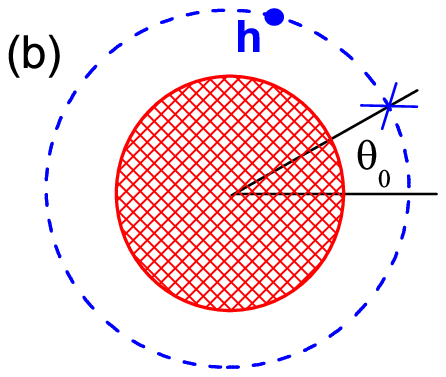}
\caption{ Schematic representation of a polarized neutral exciton
in a quantum ring (left) and in a type-II quantum dot (right) }
\label{fig:Chart}
\end{figure}

If both electrons and holes are strongly confined in the radial
direction, the corresponding dynamics is essentially
one-dimensional and can be described by two concentric rings as on
the left panel in Fig.\ \ref{fig:Chart}a (henceforth referred to
as {\it polarized quantum ring}).

On the other hand, if only the outer carrier is strongly confined
radially, then the inner carrier's wavefunction has a more
extended character and the system will have a type-II quantum dot
characteristic distribution of carriers (Fig.\ \ref{fig:Chart}b).
In this situation, the confining potential profile is such that
one of the carriers is confined inside the dot while the other is
kept spatially separated on the outside. For disk-like quantum
dots, the outside carrier is kept on a ring trajectory due to the
Coulomb attraction between the carriers. In both cases, the
polarized nature of the neutral exciton gives rise to oscillations
on the ground state energy as the magnetic flux through the ring
changes, due to the accumulation of a net Ahanonov-Bohm phase on
the electron-hole pair wavefunction.
\cite{Govorov02,UlloaPE02,GovorovPE02}

The parameters used in the remaining of the paper model the
experimental system described in the work of Ribeiro and
collaborators. \cite{Ribeiro03} In that reference, Aharonov-Bohm
oscillations on the PL energy for neutral magnetoexcitons were
reported in InP/InAs self-assembled quantum dots. The band
alignment is expected to be type-II, with the electron confined
inside the disk-like quantum dot and the hole wavefunction
localized on its outside due to the electrostatic potential
barriers at the interface (as in Fig.\ \ref{fig:Chart}b). We
should also mention that built-in strains in these QD structures
could result in an effective ring-like confinement potential for
the electron as well (as in Fig.\ \ref{fig:Chart}a).
\cite{Tadic02} As such, we will study possible qualitative
differences of the two situations and compare with experiments.

In order to compare our theoretical findings to the experimental
data, we set the effective masses $m^{*}_e=0.073 m_e$,
$m^{*}_h=0.255 m_e$, and ring radii $R_e=16$nm and $R_h=19$nm in
our calculations. Those values for $R_{e(h)}$ come from direct
imaging of the structures as well as from fitting the observed
spectral features. We also consider the experimental value for the
size dispersion in the dots: $\Delta R \approx 0.8$nm, an
important element in comparison with experiments.

For such parameters, typical single-particle confinement energy
scales are $0.1-0.5$meV. It is clear from the outset that if
charged impurities with strong trapping potentials of order $U
\sim 5$ meV were present, the corresponding wavefunction
localization would be so large that the Aharonov-Bohm oscillations
could not survive. We then consider perturbative impurity effects,
with weak potential strengths that may arise from local strain
effects due to lattice mismatches, distant charge centers and
other lattice defects near the InP/InAs interfaces.
\cite{Evaldo_priv} For concretness, we use the impurity potential
strength as $U^{\imp}_{h}=0.015$meV, and $U^{\imp}_{e}=0.023$meV
for holes and electrons, respectively, unless otherwise stated.

One should notice that even though our energy and length scales
are set for comparison with the experiment, the validity of our
results is not restricted to those parameters. In fact, as we will
see, our predictions hold as long as $U^{\imp} \ll
\hbar^2/2m^*R^2$.
%
\section{Theoretical Model}
\label{sec:theomod}

The system to be modelled displays a polarized neutral exciton
confined on a 2D ring-like geometry and subjected to a
perpendicular magnetic field. Depending on the structure, the
geometry is that of a QR or an indirect type-II dot. We present
the relevant potentials below.

In addition, we consider the effect of scattering impurities on
the ring trajectories. Although we propose a relatively simple
model for the exciton in the ring-like structure, it yields the
necessary information for a qualitative comparison with
experimental results.

\subsection{Polarized Quantum Ring}
\label{sec:Model_polQR}

We first consider the case where the confining width for {\it both
} carriers is small compared to the ring radius and include single
and multiple $\delta$-scattering impurities along the confining
region. The impurities are located on fixed angular positions
$\theta^0_i$ (which can be different for electrons and holes). The
Hamiltonian for the polarized Quantum Ring reads:
\ba
\label{eq:Hamilt}
H = \frac{\left({\bf p}_e - \frac{e}{c} {\bf A}({\bf
r}_e)\right)^2}{2m^{*}_e}   + \frac{\left({\bf p}_h + \frac{e}{c}
{\bf A}({\bf r}_h)\right)^2}{2m^{*}_h}  +  V_{\mbox{\scriptsize ring}} + \nonumber \\
+ \sum_{ij} U^{\imp}_e \delta(\theta_e - \theta^0_i ) + U^{\imp}_h
\delta(\theta_h - \theta^0_j ) + V_{eh}({\bf r}_e, {\bf r}_h)
\ea
where $m^{*}_{e(h)}$ are the electron (hole) effective masses,
${\bf A}$ is the vector potential and $U^{\imp}_{e(h)}$ are the
electron (hole) impurity strengths. The last term is the Coulomb
electron-hole interaction.

Following the discussion on section \ref{sec:refsys}, the
effective Bohr radius for our reference system will be of order
$a^{*}_B \sim 10$nm. Therefore, even though $R/a^{*}_B \sim 2$, we
have $w/a^{*}_B \ll 1$ if the radial confinement is strong ($R \gg
w$). Since our main interest resides on the impurity effects in
strongly confined non-charged excitons ($w/a^{*}_B \ll 1$), the
attractive interaction $V_{eh}$ has a perturbative effect in the
level structure and its main effect is to provide weak correlation
effects and a constant binding energy shift $E^{B}_{\exct}$, which
we can safely ignore.

\begin{figure}[h!]
\includegraphics[height=0.9\columnwidth,width=1.0\columnwidth]{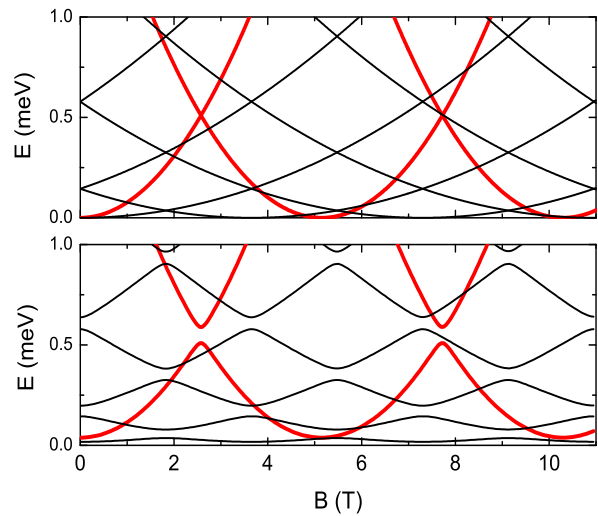}
\caption{ Electron (red thick line) and hole (black thin line)
energy levels as function of magnetic field for the clean (top)
and impurity-distorted (bottom) cases. }
\label{fig:R19}
\end{figure}

The ``clean" ($U^{\imp}_{e(h)}=0$) Hamiltonian is separable and
can be solved analytically. The angular energies and
eigenfunctions and are given by
\be
E^{(0)}_{i}(l_{e(h)}) = \frac{\hbar^2}{2m_{e(h)}R^2_{e(h)}} \left(
l_{e(h)} \mp \frac{\Phi_{e(h)}}{\Phi_0} \right)^2
\ee
\be
\varphi^{e(h)}_{i}(\theta) = \frac{1}{\sqrt{2 \pi}} e^{i l_{e(h)}
\theta} \equiv \left<{\bf r}|l_{e(h)}\right>,
\ee
where $R_{e(h)}$ and $l_{e(h)}$ are the electron (hole) confining
radius and angular momentum value, respectively, $\Phi_{e(h)}=\pi
R^2_{e(h)} B$ are the individual fluxes and $\Phi_0=h/e$ is the
unit quantum flux. The energy levels of electron and hole on the
ring are shown on the top panel of Fig.\ \ref{fig:R19}.

In this model, the excitonic states are given by a superposition
of electron and hole states $\left|\Psi_{\exct} \right>=\left|l_e
\right> \otimes \left|l_h \right> $. These states have
well-defined angular momentum values given by $L=l_h+l_e$, and
energies given by $E^{\exct}=E^{(0)}_{e}+E^{(0)}_{h}$. The ground
state angular momentum changes whenever either $\Phi_e/\Phi_0$ or
$\Phi_h/\Phi_0$ is a half-integer. In a clean system, these
angular momentum changes of the ground state result in sharp dark
$\leftrightarrow$ bright transitions in the PL of the QR.
\cite{Govorov02}

The impurities are included by numerical diagonalization of
Hamiltonian (\ref{eq:Hamilt}) in this $\{\left|\Psi_{\exct}\right>
\}$ basis. The single-particle matrix elements
$\left<l^{\prime}|U^{\imp}|l \right>$ can be calculated
analytically, giving:
\be
\left<l^{\prime}|U^{\imp}|l \right> = \left(\frac{U^{\imp}}{2 \pi}
\right) \sum_j e^{i \Delta l \theta^0_j} \, ,
\label{eq:MatrixElements}
\ee
where $\Delta l=l-l^{\prime}$, and $\theta_0$ is the angular
position of the impurity, as shown on Fig.\ \ref{fig:Chart}(b).
From Eq.\ (\ref{eq:MatrixElements}), one can readily see that the
impurity potential breaks the rotational symmetry of the system
since it couples all $\{ \left| l \right> \}$ states. Thus, the
excitonic states will no longer have definite angular momentum but
rather be a linear combination of the form:
\be
\left|\Psi^{\exct}_{k} \right>  = \sum_{l_h} C^{h}_{l_h, k}
C^{e}_{l_e, k} \left|\varphi^{h}_{l_h} \right> \otimes \left|
\varphi^{e}_{l_e} \right>
\ee

The effect of such symmetry breaking on the energy levels is seen
in Fig.\ \ref{fig:R19} in the case of a single impurity.  All the
level crossings at magnetic field values where
$\Phi_{e(h)}/\Phi_0=n/2$, present in the clean system (top panel),
become {\it anticrossings} with width proportional to
$U^{\imp}_{e(h)}$, as shown in the bottom panel. It is worth
mentioning that the spectrum remains unchanged under angular
displacements of the single impurity, since $\theta^0 \rightarrow
\theta^0 + \Delta \theta$ only gives a phase to the matrix
elements (\ref{eq:MatrixElements}).

\begin{figure}[h!]
\includegraphics[height=0.9\columnwidth,width=1.0\columnwidth]{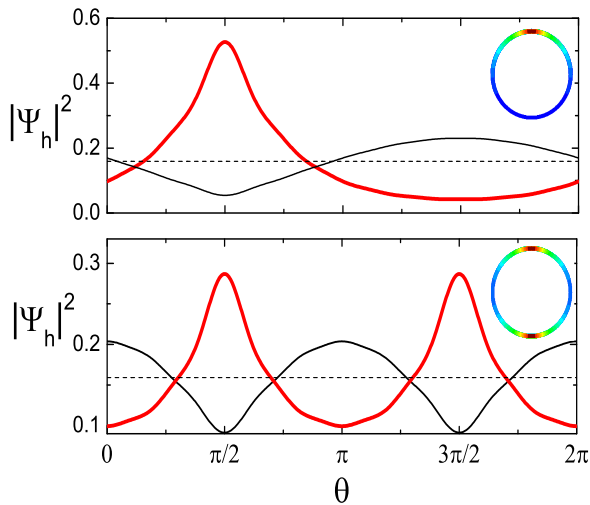}
\caption{ (color) Hole probability ground-state distribution for
attractive (red line and insets) and repulsive (black line)
impurities at $\theta_0=\pi/2$ (top panel) and two impurities at
$\theta_0=\pi/2$ and $\theta_0=3\pi/2$ (bottom panel). Dashed line
shows the uniform $(2\pi)^{-1}$ probability density for the clean
case. }
\label{fig:PsiSqr}
\end{figure}

As a consequence of impurity scattering, both the hole and
electron wavefunctions tend to be localized around $\theta^0_i$.
This effect is shown on Fig.\ \ref{fig:PsiSqr} for the hole
probability density $|\psi_h(\theta)|^2$ for one impurity, as well
as for the case of two impurities separated by a distance $\Delta
\theta$ at $B=0$. In the latter case, the wavefunction is pinned
at $\theta_0$ and $\theta_0 + \Delta \theta$ when $U^{\imp}<0$ and
is repelled from these angles for $U^{\imp}>0$. We find this
pinning to be only weakly dependent on magnetic field.
\subsection{Type-II Quantum Dots}
\label{sec:Model_tIIQD}

The Hamiltonian for the type-II ring-confinement model reads:
\ba
H = H_{\mbox{\scriptsize Dot}} + \frac{1}{2m^{*}_h} \left({\bf
p}_h + \frac{e}{c} {\bf A}({\bf r}_h)\right)^2 +
V_{\mbox{\scriptsize ring}} +
\nonumber \\
\sum_{i} U^{\imp}_h  \delta(\theta_h - \theta^0_i )  +
V_{e-h}({\bf r}_e, {\bf r}_h) \, ,
\label{eq:HamiltT2QD}
\ea
where $H_{\mbox{\scriptsize Dot}}$ describes the electron confined
in a parabolic dot with a characteristic frequency $\omega_0$ and
under the influence of a magnetic field. In the absence of
spin-orbit interactions, the electron energies are given by the
Fock-Darwin levels:
\be
E^{e}_{n l \sigma} = \left( 2n + |l| +1 \right) \hbar \Omega +
\frac{l}{2} \hbar \omega_c + g \mu_B \frac{B \sigma}{2} \, ,
\ee
where $n$ is a positive integer, $l$ is the angular momentum,
$\omega_c$ is the cyclotron frequency and
$\Omega=\sqrt{\omega^2_0+\omega^2_c/4}$ is the effective electron
frequency. The Zeeman splitting term does not alter our results
qualitatively and is disregarded in the following calculations.

We focus on impurity effects on the hole outside the QD. Notice
that (i) the effects of impurity scattering are stronger on the 1D
hole confinement as compared to the electronic 2D confinement and
(ii) the ABE reflects the phase acquired by the hole wavefunction
and would not be significatively affected by scattering processes
inside the quantum dot.

\begin{figure}[h!]
\includegraphics[height=0.9\columnwidth,width=1.0\columnwidth]{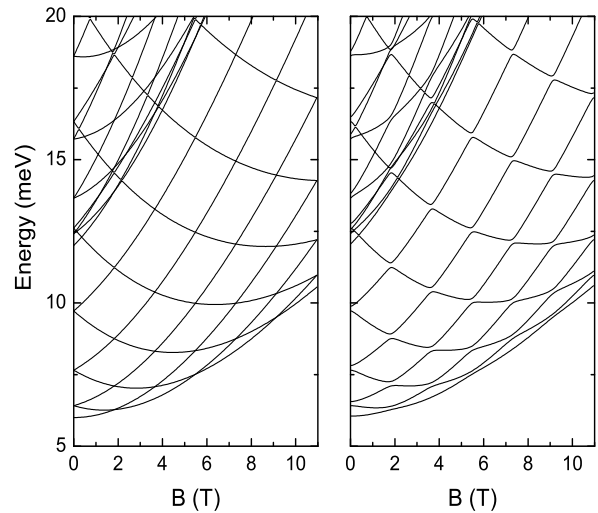}
\caption{ Excitonic energy levels as function of magnetic field on
a type-II quantum dot for the clean (left) and impurity-distorted
(right) cases. }
\label{fig:T2QDExciton}
\end{figure}

The excitonic energy levels obtained from (\ref{eq:HamiltT2QD}) in
the strong confinement regime are shown in Fig.\
\ref{fig:T2QDExciton}, for the cases with and without impurities.
The low-lying states correspond to the combination of the first
electronic states ($l_e=0$) with low-lying hole states. As in the
polarized QR case, the impurity potential induces anticrossings on
the whole spectrum, clearly seen in the right panel.
%
%
%
%
%
%
\section{PL emission intensity}
\label{sec:PLemission}

Once the spectral characteristics of the system are obtained, the
photoluminescence emission intensity can be calculated. We
consider optical interband transitions near the $\Gamma$ point of
the solid. Since the photon angular momentum is taken up by the
conduction-valence band transition matrix element, the emission
intensity is proportional to the probability of finding the
exciton on the $L=0$ state and also to the overlap between the
electron and hole wavefunctions. The optical emission occurs then
only if $l_e=-l_h$. This represents the {\it selection rules} on
the electron emission. \cite{Govorov02,GovorovPE02,UlloaPE02} If
the exciton is on state $\left| \Psi^{\exct}_i \right>$, the
emission intensity $I_i$ is then given by:
\be
I_i \propto |A_i|^2 P^{L=0}_i \, ,
\label{eq:Intensity}
\ee
with
\be
P^{L=0}_i = \left| \int \Psi^{L=0}_{i} ({\bf r},{\bf r}) d{\bf r}
\right|^2 \, ,
\ee
and where $\Psi^{L=0}_i$ is the projection of the excitonic state
in the $L=0$ state, given by
\ba
\Psi^{L=0}_{i} (\theta,\theta) = \sum_{l_h} C^{h}_{l_h, i}
C^{e}_{l_e, i} \varphi^{h}_{l_h}(\theta)
\varphi^{e}_{(l_e=-l_h)}(\theta)  \nonumber \\
\Longrightarrow \int \Psi^{L=0}_{i} (\theta,\theta) d\theta =
\sum_{l_h} C^{h}_{l_h, i} C^{e}_{l_e(=-l_h), j} \, ,
\ea

In Eq.\ \ref{eq:Intensity}, $|A_i|^2$ is the electron-hole overlap
radial integral. As discussed on section \ref{sec:Model_polQR},
the wavefunctions tend to be localized near the impurities and
such localization modifies the overlap integral $|A|^2$, and
therefore the emission intensity. However, once the wavefunctions
are localized on the angular variable, the radial overlap has only
a weak dependence with field, as long as the radial confinement is
strong. Since we are interested on how the intensity changes with
magnetic field, we take $|A|^2$ constant for simplicity since it
does not alter our field-dependent results qualitatively.

The total emission intensity at a temperature $T$ will be given by
the thermal population average over the emission from different
states:
\be
I_{PL} = \frac{\sum I_i e^{-\beta E^{\exct}_i}}{\sum e^{-\beta
E^{\exct}_i}},
\ee
where $\beta=(k_B T)^{-1}$ and $E^{\exct}_i$ is the energy of the
$i$th excitonic state.

When impurities are considered, the total angular momentum $L$ is
no longer a good quantum number since the impurity scattering
mixes the angular momentum states. Therefore, a finite emission
intensity is expected for all magnetic field values, unlike the
situation in the clean system which exhibits sharp transitions to
bright exciton states.

\subsection{Type II Quantum Dots}

\begin{figure}[h!]
\includegraphics[height=1.0\columnwidth,width=1.0\columnwidth]{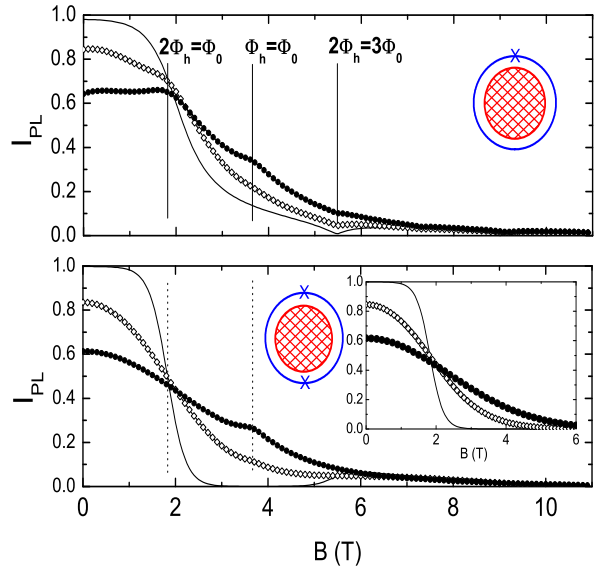}
\caption{ Photoluminescence intensity as function of magnetic
field for a electron-hole pair on a type-II QD with one (top) and
two symmetric impurities (bottom). Curves for $T=0.5$K (solid),
$T=2$K (diamonds), $T=4$K (filled circles) are shown. Vertical
lines are guides to the eye, representing $\Phi_h/\Phi_0=n/2$.
Inset: Intensity curves for the no-impurity case.}
\label{fig:IPLV01TIIQD}
\end{figure}

The plots on Fig.\ \ref{fig:IPLV01TIIQD} show the
photoluminescence intensity as function of the magnetic field for
the type-II-quantum-dot magnetoexciton. For the clean case (shown
in inset), a clear drop is seen after $\Phi_h/\Phi_0=1/2$ ($B
\approx 1.8$T for $R_h=19$nm), when the hole ground-state angular
momentum $l_h$ changes from $0$ to $1$. This is a consequence of
the emission selection rules. For low temperatures, the emission
comes mainly from the ground state and therefore the drop is more
abrupt. For higher temperatures, the $L=0$ excited states
contribute to the emission and the drop in intensity is smoother.

The presence of impurities changes this behavior qualitatively.
The cases of both single and double symmetric defects are shown on
 the top and bottom panels of Fig.\ \ref{fig:IPLV01TIIQD}, respectively. The
excitonic states are linear combinations of states with different
$L$, so that a non-zero $L=0$ component is present on the ground
state even above $\Phi_h/\Phi_0=1/2$. Therefore, the low
temperature drop in the intensity is much less abrupt than in the
clean case, as expected.

It is interesting, moreover, that at higher temperatures new PLI
features arise when $\Phi_h/\Phi_0=n/2$ ($n=1,2,3,...$). In the
single impurity case, anticrossings occur at $n=1,3,5,...$
(involving the ground state and 1st excited state), and also at
$n=2,4,...$ (between the first excited state and the second
excited state) as shown in the left panel of Fig.\
\ref{fig:T2QDExciton}. At these anticrossings, the $\left|L=0
\right>$ components of the crossing states switch. If one of the
crossing states had large $\left|L=0 \right>$ component (i.e.,
``$L=0$ character") before the crossing, it will likely have a
smaller component after the crossing. These changes in character
for the first excited states give rise to small peaks or dips on
the overall PLI at higher temperatures. At $n=1$, for instance, as
the ground state changes from  ``$L=0$ character" to ``$L=1$
character", the opposite happens to the first excited state. This
``$L=1 \rightarrow L=0$" character transition on the excited state
will give a positive second-order contribution to the intensity.
Most importantly, at $n=2$, even though there is no ground state
crossing, the second excited state has a ``$L=2 \rightarrow L=0$"
change in character, which also manifests itself as a peak on the
PLI for $T>2$K (see spectrum in bottom panel on Fig.\
\ref{fig:T2QDExciton}). On the other hand, this state's character
changes ``$L=0 \rightarrow L=3$" at $n=3$, which in turn gives a
negative contribution, seen as a sharp dip on the PLI at
$\Phi_h/\Phi_0=3/2$. These variations in PLI versus magnetic field
at well-defined multiples of the AB flux provide then unique
features that one can relate to the role of impurity/defect
potentials affecting the exciton.

For the special case of two symmetrical impurities in the ring, an
additional symmetry is introduced in the system: the coupling
given by Eq.\ (\ref{eq:MatrixElements}) vanishes whenever $\Delta
l_h$ is an odd number. In terms of symmetry, such arrangement is
equivalent to having an \textit{elongated} ring instead of a
circular one. As a consequence, the anticrossings at odd $n$
disappear while the ones at even $n$ remain. The effect on the
intensity is that a plateau near zero intensity is seen for low
temperatures between $B \approx 2$T ($\Phi_h/\Phi_0 \approx 1/2$)
and $B \approx 5$T ($\Phi_h/\Phi_0 \approx 3/2$), before the PLI
grows again for $B \gtrsim 5.5$T.

\subsection{Quantum Rings}

\begin{figure}[h!]
\includegraphics[height=0.95\columnwidth,width=1.0\columnwidth]{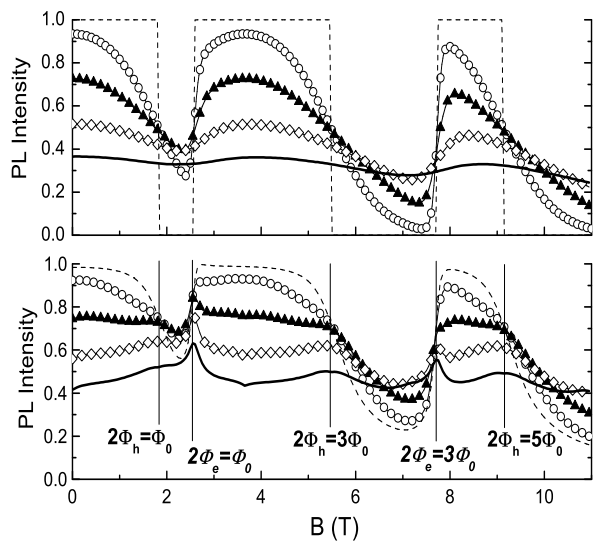}
\caption{ Photoluminescence intensity as function of magnetic
field for clean (top) and single impurity-distorted (bottom) cases
at different temperatures: $T=0$K (dashed line), $T=0.5$K
(circles), $T=1$K (triangles), $T=2$K (diamonds) and $T=4$K (thick
line). The vertical lines in the bottom panel are guides to the
eye, displaying the magnetic field values for
$\Phi_{h(e)}/\Phi_0=n/2$ . }
\label{fig:IPLV0_V01}
\end{figure}

When both the hole and the electron are confined on a ring
structure, the overall picture differs from the type-II quantum
dot case. For the clean case and at zero temperature, the PLI
displays sudden drops whenever the excitonic ground state angular
momentum goes to $L \neq 0$ states, i.e. whenever
$\Phi_{h(e)}/\Phi_0$ is a half-integer. This leads to a series of
dark and bright exciton windows in magnetic field,
\cite{Govorov02} shown as a dashed line on the top panel in Fig.\
\ref{fig:IPLV0_V01}. For higher temperatures, the thermal
occupation of higher excitonic states smoothes out the
transitions.

Impurity scattering changes this picture qualitatively in this
spectrum as well, and introduces new field-dependent features on
the PL intensity. As one impurity is added to the system, the PLI
is non-zero for all magnetic field values \textit{even at
zero-temperature}, with a mean-value larger than in the clean case
(dashed line on the bottom panel in Fig.\ \ref{fig:IPLV0_V01}).
This is a direct consequence of angular momentum mixing since the
ground state will always have an $|L=0 \rangle$ component for all
values of magnetic field.

For higher temperatures, a pronounced peak appears at
$\Phi_e/\Phi_0=1/2$ ($B \approx 2.6$T) due to a ground state
anticrossing (see Fig.\ \ref{fig:R19}). Additional peaks and
valleys can be seen for higher temperatures due to anticrossings
in the excited states, in a similar fashion as to the type-II
quantum-dot case. However, an important difference is that peaks
are seen for magnetic field values where \textit{either
$\Phi_e/\Phi_0$ or $\Phi_h/\Phi_0$} is a half-integer. The
features for electrons are sharper due to their smaller mass and
steeper level dispersions.

These findings could contribute to the understanding of recent
experimental results on photoluminescence of neutral excitons in
InP/GaAs type-II quantum dots. \cite{Ribeiro03,Evaldo_priv}
Fluctuations on the PLI are found at magnetic values which are
delayed with respect to $\Phi_h/\Phi_0=n/2$. These magnetic field
values would correspond well to changes in $\Phi_e/\Phi_0$ had the
electron been on a ring structure with a radius equal to the
estimated QD radius $R_{\mbox{\scriptsize Dot}}$. Such structure
could come from deformations in the conduction band near the dot's
edge caused by strain effects. \cite{Tadic02} Our results show
that the presence of impurities and the effective QR structure
would give rise to fluctuations at these magnetic fields.

Such features on the PL intensity could also be used to have an
experimental access into the characteristics of the disorder
potential. The size of the energy gaps in the spectrum (Fig.\
\ref{fig:R19}) are directly related to the impurity strength
$U^{\imp}$. The results presented in this section use $U^{\imp}_e
\approx 0.02$meV and $U^{\imp}_h \approx 0.04$meV, which yields
gaps of about $\Delta E_e \approx 0.08$meV for electrons and
$\Delta E_h \approx0.06$meV for holes. Although fine tuning of
those parameters is clearly possible, it is interesting that such
relatively small values lead to strong modifications of the
exciton spectra.
\begin{figure}[h!]
\includegraphics[height=0.78\columnwidth,width=1.0\columnwidth]{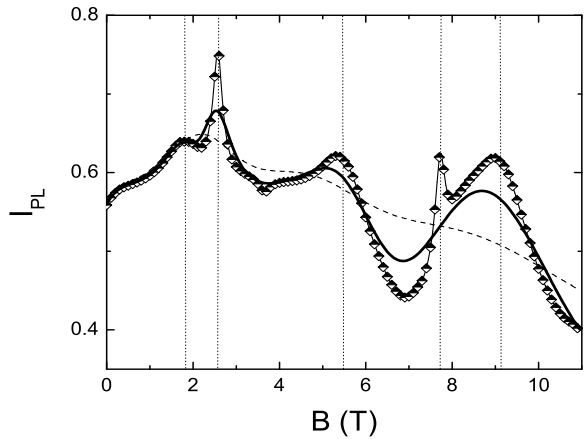}
\caption{ PL intensity as function of magnetic field for the
impurity-distorted case for $T=2$K. Diamonds denote the intensity
for fixed $R_h=19$nm and $R_e=16$nm. Thick and dashed lines denote
size-averaged intensities, with gaussian dispersions $\Delta
R=0.8$nm and $\Delta R=2$nm, respectively. }
\label{fig:IavXRV01T2}
\end{figure}

There is an additional important consideration. The results so far
consider a single polarized QR and a pertinent inquiry is what
would be the effect of the size distribution of the structures
used in experiments. We can consider these effects assuming a ring
ensemble with gaussian distributed radii (Fig.\
\ref{fig:IavXRV01T2}). As a general trend, a broad size
distribution tends to blur the impurity effects, specially at
higher magnetic fields. However, if the size distribution is as
found in experiments ($\Delta R/R_0 \approx 4\%$), these
impurity-related effects are still seen for magnetic field values
of order $\Phi/\Phi_0 \sim 1$ ($B \approx 4$T). We should
especially mention that peaks in the PLI at certain flux values
are quite robust to ensemble average, even as the dark/bright
exciton transitions are made smoother.

We summarize this discussion by saying that the systematics of
this behavior, robustness to ensemble average and temperature
effects, and even qualitative agreement with experiment,
definitively point out for an effective QR geometry of the system.

\section{Concluding Remarks}
\label{sec:final}

We have considered the effects of disorder on the Aharonov-Bohm
effect in the optical emission of type-II quantum dots and quantum
rings. As a general trend, the scattering potential breaks the
rotational symmetry, thus coupling the angular momentum states.
The effect of weak impurities does not preclude the Aharonov-Bohm
oscillations in the optical spectrum, but rather induces
additional features on the photoluminescence intensity at certain
magnetic field values. Experimental systems have routinely high
mobilities, yet some disorder is present and these PLI features
could provide additional information on the structure of the
impurity potential. They can also be used to probe the symmetries
of the quantum rings, allowing, for example, one to discern
between a circular structure and an elongated one.

Furthermore, our results could give insights on unexplained
experimental results seen on the Aharonov-Bohm effect in neutral
InP type-II quantum dot excitons. \cite{Ribeiro03} Our analysis
suggests that the unexpected magnetic field behavior of the
intensity seen in the experiment could be explained if disorder
and specific confinement of electrons and holes are taken into
account.


\acknowledgments We would like to thank Evaldo Ribeiro, Gilberto
Medeiros-Ribeiro and Mikhail Raikh for valuable conversations and
suggestions. This work was supported by FAPESP (grants 01/14276-0
and 03/03987-9), the US DOE (grant DE-FG02-91ER45334) and the
Volkswagen Foundation.

%


\end{document}